\documentclass[amsmath,amssymb,aps,prd,showpacs,nofootinbib]{revtex4}
\usepackage{graphicx}
\usepackage{amsfonts}
\usepackage{amssymb}
\usepackage{amsmath}
\usepackage{bm}
\usepackage{latexsym}

\newcommand{\beq}{\begin{eqnarray}}
\newcommand{\eeq}{\end{eqnarray}}
\newcommand{\be}{\begin{equation}}
\newcommand{\ee}{\end{equation}}
\def\la{\mathrel{\mathpalette\fun <}}
\def\ga{\mathrel{\mathpalette\fun >}}
\def\fun#1#2{\lower3.6pt\vbox{\baselineskip0pt\lineskip.9pt
\ialign{$\mathsurround=0pt#1\hfil ##\hfil$\crcr#2\crcr\sim\crcr}}}

\newcommand{{\SD}}{\rm SD}

\newcommand{\vep}{\bm p}
\newcommand{\veq}{\bm q}

\newcommand{\lan}{\langle}
\newcommand{\ran}{\rangle}

\begin{document}

\title{The $c\bar c$ interaction above threshold and the radiative
decay  $X(3872)\rightarrow J/\psi\gamma$}

\author{\firstname{A.M.}~\surname{Badalian}}
\email{badalian@itep.ru}

\author{\firstname{Yu.A.}~\surname{Simonov}}
\email{simonov@itep.ru} \affiliation{Institute of Theoretical and Experimental
Physics, Moscow, Russia}

\author{\firstname{B.L.G.}~\surname{Bakker}}
\email{b.l.g.bakker@vu.nl} \affiliation{Department of Physics and Astronomy,
Vrije Universiteit, Amsterdam, The Netherlands}

\date{\today}

\begin{abstract}
Radiative decays of $X(3872)$ are studied in single-channel
approximation (SCA) and in the coupled-channel (CC) approach,
where the decay channels $D\bar D^*$ are described with
the string breaking mechanism. In SCA the transition rate
$\tilde{\Gamma}_2=\Gamma(2\,{}^3P_1 \rightarrow \psi\gamma)=71.8$~keV
and large $\tilde{\Gamma}_1=\Gamma(2\,{}^3P_1\rightarrow
J/\psi\gamma)=85.4$~keV are obtained, giving for their ratio the value
$\tilde{R_{\psi\gamma}}=\frac{\tilde{\Gamma}_2}{\tilde{\Gamma}_1}=0.84$.
In the CC approach three factors are shown to be equally important. First,
the admixture of the $1\,{}^3P_1$ component in the normalized wave
function of $X(3872)$ due to the CC effects. Its weight $c_{\rm X}(E_{\rm
R})=0.200\pm 0.015$ is calculated. Secondly, the use of the multipole
function  $g(r)$ instead of $r$ in the overlap integrals, determining the
partial widths. Thirdly, the choice of the gluon-exchange interaction
for $X(3872)$, as well as for other states above threshold. If
for $X(3872)$ the gluon-exchange potential is taken the same as
for low-lying charmonium states, then in the CC approach $\Gamma_1=
\Gamma(X(3872)\rightarrow J/\psi\gamma) \sim 3$~keV is very small, giving
the large ratio $R_{\psi\gamma}=\frac{\mathcal{B}(X(3872)\rightarrow
\psi(2S)\gamma)}{\mathcal{B}(X(3872)\rightarrow J/\psi\gamma)}\gg
1.0$.  Arguments are presented why the gluon-exchange interaction
may be suppressed for $X(3872)$ and in this case $\Gamma_1=42.7$~keV,
$\Gamma_2= 70.5$~keV, and $R_{\psi\gamma}=1.65$ are predicted for the
minimal value $c_{\rm X}({\rm min})=0.185$, while for the maximal value
$c_{\rm X}=0.215$ we obtained $\Gamma_1=30.8$~keV, $\Gamma_2=73.2$~keV,
and $R_{\psi\gamma}=2.38$, which agrees with the LHCb data.
\end{abstract}


\maketitle

\section{Introduction}

In 2003 the Belle collaboration discovered the $X(3872)$ as a
narrow peak in the $J/\psi \pi\pi$ invariant mass distribution
in the decays $B\rightarrow J/\psi\pi\pi K$ \cite{ref.1}. Now its
characteristics, like the mass, the strict restriction on the width,
$\Gamma \la 1.2$~MeV, and the charge parity $C=+$, are well established
\cite{ref.2,ref.3,ref.4,ref.5}. In recent CDF and LHCb experiments
the quantum numbers of $X(3872)$ were determined to be $J^{PC}=1^{++}$
\cite{ref.6,ref.7}. Still, discussions about the nature of $X(3872)$
continue and to understand its exotic properties a special role is
played by the radiative decays, $X(3872)\rightarrow J/\psi \gamma$ and
$X(3872)\rightarrow \psi(3686)\gamma$, which are sensitive to the behavior
of the $X(3872)$ wave function (w.f.) at medium and large distances.

The first evidence for the decay $X(3872)\rightarrow J/\psi
\gamma$ was obtained by the Belle collaboration \cite{ref.8}
and confirmed by the BaBar collaboration \cite{ref.9}. Later
BaBar has also observed the radiative decay $X(3872)\rightarrow
\psi(3686)\gamma$ and determined the branching ratio fraction $\rm
R_{\psi\gamma}=\frac{\mathcal{B}(X(3872)\rightarrow \psi(3686)
\gamma)}{\mathcal{B}(X(3872)\rightarrow J/\psi\gamma)}=3.4\pm 1.4$
\cite{ref.10}. However, Belle has not found evidence for the radiative
decay $X(3872)\rightarrow \psi(3686)\gamma$ and put an upper limit for
the ratio \cite{ref.11},
\begin{equation}
 R_{\psi\gamma} < 2.1.
\label{eq.1}
\end{equation}
Recently the LHCb group \cite{ref.12} has observed the decay
$X(3872)\rightarrow \psi(3686)\gamma$ with a good statistics and
determined its value to be
\begin{equation}
 R_{\psi\gamma}=2.46\pm 0.64\pm 0.29,
\label{eq.2}
\end{equation}
while in Ref.~\cite{ref.13} the weighted average over three groups of
measurements was determined to be $\bar R_{\psi\gamma} = 2.31\pm 0.57$.

Unfortunately, theoretical predictions for the partial
widths $\Gamma_1$ and $\Gamma_2$ of the radiative decays
$X(3872)\rightarrow J/\psi\gamma$ and $X(3872)\rightarrow
\psi(3686)\gamma$, respectively, vary widely in different models
\cite{ref.14,ref.15,ref.16,ref.17,ref.18,ref.19,ref.20,ref.21,ref.22}
(see also the recent reviews \cite{ref.23,ref.24}). If $X(3872)$
is considered as a pure $2\,{}^3P_1$ charmonium state, then
a large value $R_{\psi\gamma}\simeq 5$ is obtained as shown in
Refs.~\cite{ref.16,ref.17,ref.18,ref.19}, exceeding the LHCb result. On
the contrary, in a molecular picture the transition $X(3872)\rightarrow
\psi(3686)\gamma$ is suppressed and the ratio $R_{\psi\gamma}$ should
be much smaller \cite{ref.14,ref.15}.  Also in many-channel models,
both in the $\,{}^3P_0$ model \cite{ref.17,ref.18} and the Cornell model
\cite{ref.19}, large values $R_{\psi\gamma}\geq 5.0$ were predicted.

It is surprising that the predicted widths $\Gamma_1$ and $\Gamma_2$
strongly differ even within the single-channel approximation (SCA),
i.e., when $X(3872)$ is supposed to be the $2\,{}^3P_1$ charmonium
state \cite{ref.16,ref.17,ref.18,ref.19}. It occurs because different
parameters in the potentials and different kinematics are used. Therefore
it remains unclear to what degree the theoretical predictions refer to
specific features of the radiative decays or to the choice of fitting
parameters and kinematics. Meanwhile, during the last decade within the
field correlator method \cite{ref.25,ref.26,ref.27} and in lattice QCD
\cite{ref.28,ref.29} the static potential was shown to be universal in
the region $r \la 1.5$~fm, where there is no creation of light $q\bar q$
pairs. Here we shall use this information and perform a parameter-free
analysis of the charmonium states with the use of the relativistic string
Hamiltonian (RSH) \cite{ref.25}.

It was already underlined by Li and Chao \cite{ref.16} and in
Ref.~\cite{ref.24} that the use of relativistic kinematics is very
important for the radiative decays and in nonrelativistic approximation
the overlap integrals, determining the partial widths, may significantly
differ from those in relativistic calculations.

In our previous paper \cite{ref.21} the rates and the ratio
$R_{\psi\gamma}$ were calculated considering $X(3872)$ (with
$J^{PC}=1^{++}$) in the coupled-channels (CC) approach, where the
coupling to the $D\bar D^*$ channels is determined by the relativistic
string-breaking mechanism \cite{ref.20,ref.21,ref.22}. This mechanism
was successfully applied to $X(3872)$, explaining it as the $2\,{}^3P_1$
charmonium state shifted down due to CC effects and appearing as a sharp
peak just at the $D^0{\bar D}^{*0}$ threshold \cite{ref.20}. Owing to such
CC effects the w.f. of $X(3872)$ acquires an admixture $c_{\rm X}(E_{\rm
R})$ from the $1\,{}^3P_1$ component, equal to $\sim 0.20$. Although this
mixing parameter is not very large, it nevertheless strongly affects the
value of the partial width $\Gamma_1$, in which the overlap integral is
small and very sensitive to different corrections.

However, in Ref.~\cite{ref.21} in the overlap integral $K_1=\lan
\varphi(2P)| g(r)|\varphi(1S)\ran$ the multipole moment $g(r)$ was
replaced by $r$, which is a common practice. In this approximation
$\tilde{K_1}=\lan \varphi(2P)|r|\varphi(1S)\ran=0.21$ has a small value,
close to those given in Refs.~\cite{ref.18,ref.24}. It appears that
the magnitude of $K_1$, defined using the multipole function $g(r)$,
is a factor of two larger than $\tilde{K_1}$, increasing the partial
width $\Gamma_1$ by about four times. Thus the replacement of $g(r)$
by $r$ cannot be applied to this radiative decay, proceeding with a
large photon energy.

We also show that the integral $K_1$ strongly depends on the
gluon-exchange (GE) part of the static potential and consider two
possibilities: First, when in $X(3872)$ the GE potential is taken the
same as for the low-lying states (below threshold). In this case the
results are different in the SCA and in the CC approach. The second
possibility implies that in higher charmonium states (above threshold),
including $X(3872)$, the GE interaction is suppressed. For such a $c\bar
c$ dynamics the partial width $\Gamma_1$ increases, while the partial
width $\Gamma_2=\Gamma(X(3872)\rightarrow \psi(3686)\gamma)$ weakly
depends on the GE interaction.

We use here the CC approach, where the admixture from the continuum to the
w.f. of $X(3872)$ is defined by the mixing parameter $c_{\rm X}(E_{\rm
R}=3.872$~GeV), which is calculated here with a good accuracy: $c_{\rm
X}=0.200\pm 0.015$. Although its value is not large, it nevertheless
gives an important contribution to $\Gamma_1$.

Also in our analysis we lay the stress on the  choice of the $c$-quark
mass, the QCD constant $\Lambda(n_f=3)$, and the string tension,
which cannot be arbitrary parameters lest the physical picture would
be distorted.

\section{The $c\bar c$ interaction below and above the threshold}

Before studing the $X(3872)$ in the many-channel approach one needs
to know its w.f. in the SCA. The rate of the electric dipole ($E1$)
transition between an initial state $i$: $n\,{}^3P_1$ and a final state
$f$: $m\,{}^3S_1$ is given by the expression \cite{ref.23,ref.24},
\begin{equation}
 \Gamma(i \stackrel{\mathrm{E1}}{\longrightarrow} f + \gamma) =
 \frac{4}{3}\,\alpha\, e_{Q}^{2}\,k_\gamma^{3}\,(2J_f+1)
 \,{\rm S}^{\rm E}_{if} \,|\mathcal{E}_{if}|^{2},
\label{eq.3}
\end{equation}
which contains the overlap integral $\mathcal{E}_{if}$ (called also
the m.e.), calculated here via the relativistic w.f. The photon
energy $k_{\gamma}$ is defined as $k_{\gamma}= \frac{M_i^2 - M_f^2}{2
M_i}$, where $M_i (M_f)$ is the mass of the initial (final) state. In
Eq.~(\ref{eq.3}) the statistical factor ${\rm S}^{\rm E}_{if}={\rm
S}^{\rm E}_{fi}$ is given by
\begin{equation}
{\rm S}^{\rm E}_{if} = \max{(l,l^{\prime})}
 \left\{\begin{array}{ccc}
         J & 1 & J^{\prime}  \\
         \l^{\prime} & s & l
        \end{array}\right\}^{2},
\label{eq.4}
\end{equation}
which for the $E1$ transitions between the $n\,{}^3P_J$ and
$m\,{}^3S_1~(m\,{}^3D_1)$ states, with the same spin $S=1$, is equal
to $\frac{1}{9}~(\frac{1}{18})$.

The overlap integral $\mathcal{E}_{if}$, given by
\begin{equation}
 \mathcal{E}_{if} = \int dr r^2 R_{n_i,l_i}(r)\, g (r) \,
 R_{n_f,l_f}(r)= \langle n_i,l_i|g(r)|n_f,l_f \rangle,
\label{eq.5}
\end{equation}
is determined by the relativistic radial w.f. $R_{n_i,l_i}(r)$ and
$R_{n_f,l_f}(r)$. The function $g(r)$  \cite{ref.24} is:
\begin{equation}
 g(r) = \frac{3}{k_{\gamma}} [ y j_0(y) - j_1(y)]=
 \frac{3}{2} \frac{r}{y}
 \left\{\sin y \left(1-\frac{1}{y^2}\right) +\frac{1}{y}\cos y\right\},
\label{eq.6}
\end{equation}
where $j_n(y)~(n=1,2)$ are the spherical Bessel functions and the variable
$y=k_{\gamma} r/2.$ It can be easily shown that the replacement of $g(r)$
by $r$ ($g(r)< r$), used in many analyses, is a good approximation for
$y\leq 1.0$, while in the range, $1.1\leq y\leq 1.3$, the difference
$g(r)-r$ can already reach $\leq 30\%$. The zero of the function
$g(r)$ occurs at $y_0=2.74$, corresponding to very large $r$, $r\ga 10
$~GeV$^{-1}$, even for the photon energy, $k_{\gamma}=0.697$~GeV for
the $X(3872)\rightarrow J/\psi\gamma$ radiative decay. In $r$-space the
difference $g(r)-r$ remains small for photon energies $k_{\gamma}\la
0.40$~GeV.

The overlap integrals, involving the $1P$ and $2P$ states, are
denoted below as
\begin{eqnarray}
 I_m \equiv I_{1P,mS} & = & \int dr r^2 R_{1P}(r) g(r) R_{mS}(r),~(m=1,2),
\nonumber \\
 K_m \equiv K_{2P,mS} & = & \int dr r^2 R_{2P}(r) g(r) R_{mS}(r),~(m=1,2).
\label{eq.7}
\end{eqnarray}
In Eq.~(\ref{eq.7}) the relativistic radial w.f. are calculated with
the use of the RSH $H_0$, derived in Ref.~\cite{ref.25} and applied many
times to different effects in quarkonia \cite{ref.30,ref.31}.
For heavy quarkonia $H_0$ has a very simple form:
\begin{equation}
 H_0=2\sqrt{\vep^2+m_c^2} + V_{\rm B}(r),
\label{eq.8}
\end{equation}
where its kinetic term is similar to that in the relativized quark
model (RQM) \cite{ref.32}. However, in $H_0$, by derivation, the
fundamental value of the $c$-quark mass is equal to its pole mass,
$m_c\simeq 1.42$~GeV. (It takes into account corrections perturbative in
$\alpha_s(m_c)$ and corresponds to the conventional current mass $\bar
m_c(\bar m_c)\sim 1.23$~GeV \cite{ref.33}). In the RQM the $c$-quark mass
is considered as a fitting parameter, e.g., $m_c=1.628$~GeV is used in
Ref.~\cite{ref.32} and $m_c=1.562$~GeV in Ref.~\cite{ref.17}.

In the RSH the universal static potential $V_{\rm B}(r)$ contains the
confining and GE terms,
\begin{equation}
 V_{\rm B}(r)=\sigma r - \frac{4\alpha_{\rm B}(r)}{3 r},
\label{eq.9}
\end{equation}
where the string tension $\sigma=0.18$~GeV$^2$ is fixed by the slope
of the Regge trajectories for light mesons and therefore cannot be
considered as a fitting parameter.

An important step is to make the correct choice of the vector
coupling $\alpha_{\rm B}(r)$ \cite{ref.34}. To this
end, the parameters defining the coupling, are here taken in
correspondence to the existing data for the strong coupling
$\alpha_s(r)$ in the $\overline{MS}$ scheme in perturbative QCD
\cite{ref.35}. The asymptotic-freedom behavior of $\alpha_{\rm B}$
is determined by the ``vector" QCD constant $\Lambda_{\rm B}$, which,
however, is not a new parameter but expressed through
$\Lambda_{\overline{MS}}$. In particular, for $n_f=3$ the relation
$\Lambda_{\rm B}(n_f=3)=1.4753~\Lambda_{\overline {MS}}(n_f=3)$ is
valid. In pQCD
\begin{equation}
\Lambda_{\overline{MS}}(n_f=3)=339\pm 10 ~{\rm MeV}, 
\label{eq.10}
\end{equation}
was extracted with the use of the matching procedure
at the quark-mass thresholds \cite{ref.35}. This value
$\Lambda_{\overline{MS}}(n_f=3)=(339\pm 10)$~MeV gives a rather
large value of $\Lambda_{\rm B}(n_f=3)=(500\pm 15)$~MeV, while in
Ref.~\cite{ref.34} from the analysis of the bottomonium spectrum,
a smaller $\Lambda_{\rm B}(n_f=3)= (465\pm 20)$~MeV was shown to be
preferable; its value corresponds to $\Lambda_{\overline{MS}}(n_f=3)
=(315\pm 14)$~MeV, in agreement with the lower limit in Eq.~(\ref{eq.10}).

Also in Ref.~\cite{ref.34} the infrared regulator $M_{\rm B}$ was
introduced, since for $n_f=3$ the characteristic momenta $q^2$ are
rather small and nonperturbative effects become important. As shown in
Ref.~\cite{ref.36}, the regulator $M_{\rm B}$ is not a new parameter,
but expressed via the string tension $\sigma$, according to the relation
$M_{\rm B}^2= 2\pi\sigma$, derived with an accuracy $\sim 10\%$. Then
for $\sigma=0.18$~GeV$^2$ one finds that $M_{\rm B}=1.06\pm 0.10$~GeV;
here the value $M_{\rm B}=1.10$~GeV is used.

Thus the potential $V_{\rm B}(r)$ contains only fundamental quantities:
the current (pole) $c$-quark mass, the string tension from the Regge
trajectories, and $\Lambda_{\overline{MS}}(n_f=3)$.  In our calculations
the following set of parameters is used:
\begin{equation}
m_c=1.425~{\rm GeV},~~ \sigma=0.18~{\rm GeV}^2,~~ M_{\rm B}=1.10~{\rm GeV},~~
\Lambda_{\rm B}(n_f=3)=465~{\rm MeV}.
\label{eq.11}
\end{equation}
For these values of $\Lambda_{\rm B}(n_f=3)$ and $M_{\rm B}$ we find
the frozen (asymptotic) value of the coupling $\alpha_{\rm crit}=0.6086$.

A remarkable property of the RSH is that in heavy quarkonium the centroid
masses $M_{\rm cog}(nl)$ of the $nl$ multiplet exactly coincide with
the eigenvalues of the spinless Salpeter equation (SSE):
\begin{equation}
 H_0\varphi_{nl}=M_0(nl)\varphi_{nl}.
\label{eq.12}
\end{equation}
In the charmonium spectrum not many levels are below the open charm
threshold, where the universal static potential $V_{\rm B}(r)$
may be tested. However, above the $D\bar D$ threshold the $c\bar c$
interaction may differ from that for low-lying states, due to open
channels or light-quark pair creation. In the string-breaking picture it
is manifested as a flattening of the static potential at relatively large
distances and this phenomenon is seen in lattice QCD \cite{ref.37} and
also observed in the radial excitations of light mesons \cite{ref.38}.
For that reason for higher states (above threshold) we consider two
possibilities: the first one, when the GE potential is taken the same
as in Eq.~(\ref{eq.9}) and the  second case, when the GE interaction
is supposed to be suppressed for higher states, so that the $c\bar c$
dynamics is totally determined by the confining potential.  There are
several arguments in favor of the latter assumption:

\begin{enumerate}

\item Suppression of the GE interaction follows from the analysis of
the orbital Regge trajectories in charmonium  \cite{ref.39,ref.40}, which are not compatible with a significant GE
contribution to the masses of higher states.

\item The fine-structure effects, caused by the GE interaction, are
not seen for the $2P$ charmonium multiplet. For example, the measured
mass difference $\delta M(2P)= M(\chi_{c2}(2P))-M(\chi_{c0}(2P))\leq
10$~MeV, is very small compared to the same mass difference for the $1P$
multiplet: $\delta M(1P)=M(\chi_{c2}(1P)) -M(\chi_{c0}(1P))= 142$~MeV
\cite{ref.33}, being smaller by an order of magnitude. It is difficult
to explain such a suppression of the fine-structure effects, if a
universal GE potential is used both for high- and low-lying states. (The
suppression of the fine-structure effects does not contradict possible
hyperfine splittings of the higher $S-$wave charmonium states due to
a short-range hyperfine potential \cite{ref.27}).

\item Our calculations of the charmonium spectrum show that the
masses of higher states are practically the same for the linear and
linear+GE potentials, if the correct choice of the $c$-quark mass is
made (see below Table~\ref{tab.01}).
\end{enumerate}

The confining potential used
\begin{equation}
 V_{\rm C}(r) = \sigma r,
\label{eq.13}
\end{equation}
has the same string tension $\sigma=0.18$~GeV$^2 $. However, for the
linear potential the $c$-quark mass may be different, since for a totally
suppressed GE interaction there are no perturbative corrections to the
$c$-quark  mass and in this case $m_c$ is equal to the current mass
(here $\ m_c(\bar m_c)=(1.27-1.29)$~GeV is used).

In Table~\ref{tab.01} we give the centroid masses $M_{\rm cog}(nl)$
for both potentials, Eqs.~(\ref{eq.7}) and (\ref{eq.9}).
\begin{center}
\begin{table}[ht]
\caption{The centroid masses $M_{\rm cog}(nl)$ (in MeV) of higher
multiplets for the potentials $V_{\rm B}(r)$ and $V_{\rm C}(r)$
\label{tab.01}}
\begin{tabular}{|c|c|c|c|}\hline
 ~State~&~Potential $V_B(r)$~&~Potential $V_C(r)$~&~experiment~\\
\hline
         &  $~m_c=1.425$~GeV~&      $~m_c=1.290$~GeV~ &  \\
\hline

    3S  &    4092       &         4112    &      $  4039 \pm 1$\\
\hline

    4S     & 4447       &         4448    &        $4424\pm 4$\\
\hline

    2P    &  3949       &         3949    &        $3927\pm 3^{a)}$\\
\hline

    3P     & 4319       &         4298    &        absent \\
\hline

    1D     & 3802       &         3788    &        $3773\pm 3 $\\
\hline

    2D      &4187       &         4155    &        $4153\pm 3$ \\
\hline
\end{tabular}

$^{a)} $ For $M_{\rm cog}(2P)$ the experimental mass
$M(\chi_{c2}(2\,^3P_2)$ is given.

\end{table}
\end{center}

From Table~\ref{tab.01} one can see that both spectra coincide within
$\la 20$~MeV accuracy. Moreover, the masses $M(1D)$, $M(2D)$, $M(2P)$,
and $M(4S)$, calculated in SCA, are close to the experimental masses
of $\psi(3770)$, $\psi(4160)$, $\chi_{c2}(2P)$, and $\psi(4420)$,
respectively, and one can expect that their hadronic shifts are not
large, $\la 20$~MeV. It is not so for the $3S$ state, where $M(3S)$
is about $(60\pm 10)$~MeV larger than $M(\psi(4040))$, indicating the
importance of the CC effects for $\psi(4040)$.

Thus our analysis of the charmonium spectrum does not allow to
draw a definite conclusion what type of the $c\bar c$ interaction
is preferable above threshold. Therefore other phenomena in
charmonium have to be studied. One of them is just the radiative
decay $X(3872)\rightarrow J/\psi\gamma $, which, as was already
underlined by E.~Swanson \cite{ref.15}, ``is particularly
sensitive to model details".

\section{The radiative decays of $X(3872)$ in single-channel approximation}
\label{sec.III}

Here we calculate the partial widths of the radiative transitions,
$X(2\,{}^3P_1)\rightarrow J/\psi\gamma$ and $X(2\,{}^3P_1)\rightarrow
\psi(2S)\gamma$ in SCA, i.e., assuming that $X(3872)$ is the $2\,{}^3P_1$
state. Then the widths (in units GeV) can be written as

\begin{eqnarray}
 \Gamma_1^0 & = & 1.4418 ~10^{-3}~k_{1\gamma}^3 ~ |K_1|^2,
\\ \nonumber
 \Gamma_2^0 & = & 1.4418 ~10^{-3}~k_{2\gamma}^3 ~|K_2|^2,
\label{eq.14}
\end{eqnarray}
where the number $1.4418~10^{-3}$ is the product of known factors in
Eq.~(\ref{eq.3}), $k_{1\gamma}=0.6974$~GeV, $k_{2\gamma}=0.1815$~GeV,
and for $X(3872)$ as the $2\,{}^3P_1$ charmonium state, the integrals  $K_1,~
K_2$ (in the units GeV${}^{-1}$) are defined in Eq.~(\ref{eq.7}).

From Eq.~(\ref{eq.7}) and for the potential $V_{\rm B}(r)$ we obtained
\begin{equation}
 K_1 = K_{2P,1S}= - 0.418~{\rm GeV}^{-1} ,
\label{eq.15}
\end{equation}
which in SCA gives a large partial width:
\begin{equation}
\Gamma_1=85.4~{\rm keV}, 
\label{eq.16}
\end{equation}
This integral $K_1$, defined with $g(r)$, has a magnitude that
is two times larger than $\tilde {K}_1=\lan R_{2P}|r|R_{1S}\ran$,
where $g(r)$ is approximated by $r$:
\begin{equation}
 \tilde{K}_1 = - 0.210 ~{\rm GeV^{-1}},\quad \tilde{\Gamma}_1 = 21.6~{\rm keV}.
\label{eq.17}
\end{equation}
Notice, that even smaller magnitudes of $\tilde {K_1}$ were obtained in
other models \cite{ref.17}, \cite{ref.21} (see Table~\ref{tab.02}). (The
choice of the negative sign of $K_1$ will be explained in
Sect.~\ref{sec.V}). Thus, in the overlap integral determining the
radiative decay $X(2\,{}^3P_1)\rightarrow J/\psi\gamma$ with rather
large photon energy, $k_{\gamma}=0.697$~GeV, the replacement of $g(r)$
by $r$ has a poor accuracy.  On the contrary, the differences between
the overlap integral $K_2$ and the width $\Gamma_2$,
\begin{equation}
 K_2=\lan 2P|g(r)|2S\ran=2.886~{\rm  GeV}^{-1},\quad \Gamma_2=71.8~{\rm keV},
\label{eq.18}
\end{equation}
and the approximate quantities $\tilde{K}_2$ and $\tilde{\Gamma}_2$
\begin{equation}
 \tilde {K}_2 =\langle 2P|r|2S\rangle =
 3.017~{\rm GeV}^{-1},\quad \tilde{\Gamma}_2=78.5~{\rm keV}
\label{eq.19}
\end{equation}
are rather small, $\la 8\%$. Then, because of  the small value of
$\tilde{\Gamma}_1$, the ratio of the partial widths $\tilde{\Gamma}_2$ to
$\tilde{\Gamma}_1$ is large,
\begin{equation}
 \tilde {R}_{\psi\gamma} = 3.6.
\label{eq.20}
\end{equation}
On the contrary, if the integrals $K_1$ and $K_2$ are calculated with
$g(r)$: $\Gamma_1=85.4$~keV, $\Gamma_2=71.8$~keV, then
\begin{equation}
 R_{\psi\gamma} = 0.84,
\label{eq.21}
\end{equation}
is four times smaller. This ratio is not compatible with the BaBar
\cite{ref.9} and the LHCb results, Eqs.~(\ref{eq.2}), but does not
contradict the Belle restriction, $R_{\psi\gamma} \la 2.1$.

In Table~\ref{tab.02} we give the partial widths $\Gamma_1$ and
$\Gamma_2$, predicted in several relativistic models. One can see that
in the SCA, even with relativistic kinematics, the predicted values of
$\Gamma_1$ vary in a wide range, $(11-85)$~keV.

Surprisingly, the m.e. $K_1$ in Eq.~(\ref{eq.15}) and $\Gamma_1$
in Eq.~(\ref{eq.16}) calculated here with a relativistic
Hamiltonian, practically coincide with the approximate m.e. $\lan
2P|r|1S\ran=0.416$~GeV$^{-1}$ and $\Gamma_1=84.6$~keV, calculated
in a nonrelativistic model with the logarithmic potential in
Ref.~\cite{ref.42}; the predicted value found there $R_{\psi\gamma}=1.1$
is also close to our number in Eq.~(\ref{eq.21}).

\begin{center}
\begin{table}
\caption{The partial widths of the $X(3872)$ radiative decays (in keV) in
relativistic models $^{a)}$\label{tab.02}}
\begin{tabular} {|c|c|c|c|c|c|c|c|}
\hline
 ~Transition~ & $k_{\gamma}$ &[41]&[16] & [17] & [18] &~our paper~&~our paper~\\
\hline
~ Method ~   & ~ in GeV ~     &  ~  SCA~ &~  CCM~ &~    SCA~  & ~ SCA~ & ~  SCA~ & ~   CCM ~ \\
~$2^3P_1\rightarrow 1^3S_1 + \gamma$~& 0.697 & 33 & 45 & 11 & 11 & 85.4 & $37\pm 6$\\
~$2^3P_1\rightarrow 2^3S_1 + \gamma$~& 0.181 & 146 & 60 & 70 & 63.9 & 71.8 & $72\pm 1$\\
~$2^3P_1\rightarrow 1^3D_1 + \gamma$~&  0. 098 & 7.9 &  & 4.0 & 3.7 & 5.8 & \\
\hline
\end{tabular}

$^{a)}$ The partial widths are calculated either in single-channel
approximation (SCA) or within a coupled-channels method (CCM).
\end{table}
\end{center}

As a test of the w.f., defined by the RSH and used here, we
also calculated the partial widths of the $E1$ transitions:
$\chi_{cJ}(1P)\rightarrow J/\psi\gamma~(J=0,1,2)$, where all three
states lie far below the $D\bar D$ threshold and are described by the
linear+GE potential.  For $\chi_{c1}(1P)\rightarrow J/\psi\gamma$ and
$\chi_{c2}(1P)\rightarrow J/\psi\gamma$ the calculated widths appeared to
be in precise agreement with the experimental data. For these radiative
decays the calculated integral $I_{1P,1S}\equiv I(1\,{}^3P_{cJ}, 1S)
= 1.849$~GeV$^{-1}$ is taken the same for $J=0,1,2$. For $\psi(3686)$
its w.f. contains an admixture from the $1D$ state \cite{ref.43} and
therefore this radiative decay cannot be used as a test.

\begin{table}
\begin{center}
\caption{ The overlap integrals and the partial widths of the radiative decays
$\chi_{cJ}(1P)\rightarrow J/\psi\gamma$.\label{tab.03}}
\begin{tabular} {|c|c|c|c|r|r|}\hline
 Transition &~$k_{\gamma}$ (GeV)~ & ~$\langle r\rangle $ (GeV$^{-1})$~
 &~$ \langle g(r)\rangle^{a)}$  (GeV$^{-1})$&
 $\Gamma({\rm th})$  & $\Gamma({\rm exp})$\\
\hline ~~$1\,{}^3P_{c2} \to J/\psi + \gamma$ ~ & 0.429  & 1.976  & 1.849 & 390
~ ~&
$ 386\pm 17$\\
\hline ~~$1\,{}^3P_{c1} \to J/\psi + \gamma$ ~   &  0.389  & 1.976   &   1.849
&  290~ ~&
 $ 296\pm 13$\\
\hline ~~$1\,{}^3P_{c0} \to J/\psi + \gamma$ ~ & 0.303  &  1.976   &   1.849  &
137 ~ ~&
$ 133\pm 10$\\
\hline
\end{tabular}

$^{a)}$ Calculations refer to the variant with $m_c=1.425$~GeV, $\Lambda_{\rm
B}(n_f=3)=465$~MeV, and $M_{\rm B}=1.1$~GeV.
\end{center}
\end{table}

For all three radiative transitions the photon energies are not large,
$k_{\gamma}\la 0.40$~GeV, and therefore the differences between the m.e.
$\langle g(r) \rangle$ and $\langle r \rangle$ are rather small, $\la
7\% $. Nevertheless, with the use of the function $g(r)$ the partial
widths of those radiative decays decrease by $\sim 10\%$, improving the
agreement with experiment.

From Table~\ref{tab.03} one can see that the calculated values
$\Gamma(\chi_{c0}(1P)\rightarrow J/\psi\gamma)= 137$~keV,
$\Gamma(\chi_{c1}(1P)\rightarrow J/\psi\gamma) = 290$~ keV and
$\Gamma(\chi_{c2}(1P)\rightarrow J/\psi\gamma) = 390$~keV are in precise
agreement with the experimental widths: $\Gamma_{\rm exp}(\chi_{c0}\rightarrow
J/\psi\gamma)=133\pm 10$~keV, $\Gamma_{\rm exp}(\chi_{c1}\rightarrow
J/\psi\gamma )=296\pm 13$~keV, and $\Gamma_{\rm exp}(\chi_{c2}\rightarrow
J/\psi\gamma)=386\pm 17$~keV \cite{ref.33}.

\section{Dynamics of coupled channels and an admixture of neighboring states}
\label{sec.IV}

In the CC approach $X(3872)$ is treated as the original $2\,{}^3P_1~(Q\bar
Q)$ state, shifted down by the $CC$ interaction with the $D\bar D^*
(\bar D D^*)$ channel. Moreover, in Ref.~\cite{ref.20} explicit parameters
of this interaction were found, which put the pole of $X(3872)$ in the
multi-channel Green's function exactly at the position of the open channel
threshold. It is also important that within our approach the parameters
are calculated and found to lie in a narrow range, i.e., they are not
fitted. Below we use these fixed parameters to predict the admixture
generated by the $CC$ interaction, to the original $2^3P_1$ state.

Leaving the details of the $CC$ formalism to the Appendix and the
original papers \cite{ref.20,ref.21,ref.22}, we recapitulate briefly
the essence of the mixing problem.  We define the extra (open)-channel
interaction by the function $V_{CC}$ $(\veq, \veq', E)$ in momentum
space, which depends on the total energy $E$ and contains all thresholds
of the open channels. Then we are writing the full Green's function
$G_{Q\bar Q}^{(BB)}$, which describes the evolution of a $Q\bar Q$
pair (produced in the $e^+e^-$ annihilation or the $B$ decay process),
interacting in the intermediate states with $D\bar D^* (\bar D D^*)$
and finally ending up again as a $Q\bar Q$ pair (in a similar $e^+ e^-$
annihilation or the $B$ decay process), as

\begin{equation}
 G_{Q\bar Q}^{(BB)} = \sum_{n,m} (\hat B \psi_n)
 \frac{1}{\hat E - E+ \hat w} (\hat B \psi_m)^\dagger,
\label{eq.22}
\end{equation}
where the matrices are defined as $\hat E = E_n \delta_{nm}$, and
$\hat{w} = w_{nm}$ with
\begin{equation}
 w_{nm} (E) = \int \psi^\dagger_n (\veq) V_{cc} (\veq, \veq' , E) \psi_m
 (\veq') \frac{d^3\veq}{(2\pi)^3} \frac{d^3\veq'}{(2\pi)^2}.
\label{eq.23}
\end{equation}
Here the symbol $\hat B$ in $(\hat B\psi_n)$ stands for an operator,
defining the production (annihilation) process. In the case when the
pure $Q\bar Q$ state is already formed, one can put $\hat B\equiv 1$,
while in the $e^+e^-$ annihilation process the operator $\hat B$ is
proportional to $\gamma_{\mu}$.

Note, that $G^{(BB)}_{Q \bar Q}$ contains all CC contributions in the
intermediate states, but the experimental conditions require these
contributions to be projected onto the $Q\bar Q$ states. We now include
in $G_{Q\bar Q}^{(BB)}$ the possibility to emit any particle $(\gamma,
\pi, ...)$ in the intermediate state. As shown in Ref.~ \cite{ref.21}
(see Appendix), this can be done inserting the self-energy part with the
photon loop to $\hat E$ and $\hat w$:  $\hat E+\hat w \to \hat E(\gamma)
+ \hat w(\gamma)$, so that the probability of this emission is given by
the discontinuity of the self-energy part of $G_{Q\bar Q}^{(BB)}$, e.g.
\begin{eqnarray}
 Y_\gamma (E) &  = & \frac{1}{2} \Delta G_{Q\bar Q}^{(BB)}
\nonumber \\
 & = &
 \sum_{n,m,l}(\hat B \psi_n)
 \left( \frac{1}{\hat E- E+\hat w}\right)_{nm}
 \frac{1}{2i} \Delta (\hat E(\gamma)+\hat w(\gamma))_{ml}\left( \frac{1}{\hat E- E+\hat
w}\right)_{lq}(\hat B\psi_q)^\dagger
\nonumber \\
 &  =  & \sum_{k,n} a_k \psi_k \frac{1}{2i} \Delta
(\hat E + \hat w)_{kn} \psi^\dagger_n a_n^\dagger.
\label{eq.24}
\end{eqnarray}
In this way one can start in $e^+e^-$, or the $B$ decay, with the creation
of any combination $\{\hat B \psi_n\}$, which finally, at the moment of
the $\gamma$ detection, becomes the set of the states $(\psi_k + \gamma)$
with the probability amplitude $a_k(\hat B \psi_n)$.

The amplitude $a_k$ is given by
\begin{equation}
 a_ k = \sum_{l} \left(\frac{1}{\hat E- E + \hat w}\right)_{kl} (B\psi_l).
\label{eq.25}
\end{equation}
Since $\hat E$ is diagonal, $\hat E_{ik}= E_i \delta_{ik}$, all
matrix structure is due to $\hat w$. This matrix was calculated in
Ref.~\cite{ref.21} for the $1^{++}$ states, giving the values shown in
the Table~\ref{tab.04} in the Appendix. here one can see, that only
two $1^{++}$ states are strongly connected by $w_{nm}$, namely  $1^3P_1$
and $2^3P_1$, while higher states yield very small admixtures. 

Defining the amplitudes $a_1$ and $a_2$ for the $1^3P_1$ and $2^3P_1$
states, respectively, one has
\begin{equation} \frac{a_1}{a_2} = \frac{(E_2+w_{22}-E)
 \frac{B\psi_1}{B\psi_2}-w_{21}}{E_1+w_{11}-E-w_{12}
 \frac{B\psi_1}{B\psi_2}} .
\label{eq.26}
\end{equation}

With the use of the RSH we have $E_2 =3949\pm 10$~MeV,
$E_1=3523\pm 10$~MeV and then from Table~\ref{tab.03}
\begin{equation}
 E_2 + w_{22} = (3850\pm 10){\rm MeV}, \quad
 E_1+w_{11} = (3203 \pm 10){\rm MeV}.
\label{eq.27}
\end{equation}
From Eq.~(\ref{eq.26}) one can see that the solution for $a_1/a_2$
is sensitive to the original value of $E_2$, but depends weakly on
$B\psi_1/B\psi_2$ and we take for this ratio the same value as for
the $e^+e^-$  process,  namely  $|\psi_1(0)|^2/|\psi_2(0)|^2\approx
0.85$. Inserting these numbers into Eq.~(\ref{eq.26}) and taking
$E_2=3949$~MeV from Table ~\ref{tab.01} (the contribution from the
$D^*\bar D^*$ channel is neglected) one obtains the value of $c_{\rm X}$
at the resonance energy, $E_{\rm R}=3872$~MeV ($B\psi_1/B\psi_2=0$):
\begin{equation}
 c_{\rm X} (E_{\rm R})=\frac{a_1}{a_2} (E=3872~{\rm MeV}) =0.183.~
\label{eq.28}
\end{equation}

However, this number can be considered as a lower limit, since we have
not accounted for the contribution of the closed $D^*\bar D^*$ channel
to the original levels $E_1$ and $E_2$. The estimate of these $D^*
\bar D^*$ contributions to the $2^{++}(2^3P_2)$ and $0^{++}(2^3P_0)$
states was made in Ref.~\cite{ref.20} for the same CC parameters as for
$X(3872)$. It yields a shift around $-30$ MeV, in good agreement with
the $Z(3930)$ as the $2^3P_2$ state. In this case in Eq.~(\ref{eq.28})
$E_2=3920$~MeV and a larger value of $c_{\rm X}$ is obtained,
\begin{equation}
 c_{\rm X}=\frac{a_1}{a_2} (E=3872~{\rm  MeV})  \approx 0.215
\label{eq.29}
\end{equation}
As a result, we chose to accept the value of $c_{\rm X}$ from the range,
defined by Eqs.~(\ref{eq.28}) and (\ref{eq.29})
\begin{equation}
 c_{\rm X}=0.200\pm 0.015 .
\label{eq.30}
\end{equation}

\section{The coupled-channels corrections to the $X(3872)$ wave function}
\label{sec.V}

In the previous section the effects from open $D\bar D^*$ channels
were calculated, resulting in the contribution from the
$1\,{}^3P_1$ state to the normalized w.f. of $X(3872)$, which
can be represented as

\begin{equation}
 \varphi(X(3872)) = \sqrt{1-c_{\rm X}^2}~\varphi(2\,{}^3P_1) + c_{\rm X}~\varphi(1P\,{}^3P_1).
\label{eq.31}
\end{equation}
As seen from Eq.~(\ref{eq.30}) the mixing parameter $c_{\rm X}$ at the
energy $E_{\rm R} = 3.782$~GeV lies in a narrow range; this admixture
was not fitted, but calculated within the string breaking mechanism.

Moreover, in the many-channels mechanism the sign of the continuum component
cannot be arbitrary anymore and here we choose positive $c_{\rm
X}$: it takes place if the radial w.f. $R_{2P}(r)$ is chosen to be
negative at small distances (for $r\la 0.5$~fm), while the radial
w.f. of the ground state $R_{1P}(r)$ is positive in the whole
region. Just for this choice a negative value for the overlap integral in
Eqs.~(\ref{eq.15}, \ref{eq.17}) was obtained.

Due to the CC admixture in the $X(3872)$ w.f., the m.e. in the
radiative decays of $X(3872)$ can be presented as a superposition,
\begin{eqnarray}
\lan X(3872)|g(r)| J/\psi\ran & = &
  \sqrt{1-c_{\rm X}^2}~ K_1 + c_{\rm X} I_1,
\nonumber \\
\lan X(3872)|g(r)|\psi(2S)\ran & = &
 \sqrt{1-c_{\rm X}^2} ~ K_2 + c_{\rm X} I_2,
\label{eq.32}
\end{eqnarray}
where $I_n$ and $K_n~(n=1,2)$ are defined in Eq.~(\ref{eq.7}) and the radial w.f. of  $\psi(3686)$
is taken as $R(2S)(r)$, i.e., without mixing with the $1D$ state.

For $X(3872)$ two different $c\bar c$ potentials are used below.
In first case the w.f. of all states involved in the radiative
decays of $X(3872)$, are defined by the RSH with the linear+GE
potential, $V_{\rm B}(r)$, given in Eq.~(\ref{eq.9}). Then in the
m.e. Eq.~(\ref{eq.32}) the two terms have different signs and
therefore the magnitude of this m.e. is smaller as compared to the
number given in Eq.~(\ref{eq.15}). Due to this cancellation this m.e.
appears to be very small, even for the minimal value of $c_{\rm
X}({\rm min})=0.185$ from Eq.~(\ref{eq.28}):
\begin{equation}
  |\lan X(3872, 2\,{}^3P_1))|g(r)|J/\psi\ran|\leq 0.08~{\rm GeV}^{-1},
\label{eq.33}
\end{equation}
giving a small partial width $\Gamma_1=3.0$~keV.

On the contrary, for the transition $X(3872)\rightarrow \psi(2S)\gamma$
its m.e. increases to 3.30~GeV$^{-1}$ and $\Gamma_2=94$~keV is large.
Thus, within the CC approach, taking the linear+GE potential for
$X(3872)$, one obtains a small partial width $\Gamma_1\la 3$~keV and
very large value for the ratio $R_{\psi\gamma}\gg 1.0$.

The situation is different, if we assume that the GE interaction is
totally suppressed for higher charmonium states, i.e., the $c\bar c$
dynamics is determined by the confining potential. (Note that such a
suppression does not exclude the use of the hyperfine interaction, having
a very small range \cite{ref.28}, for the higher $S$-wave charmonium
states, like $\psi(4040)$, concentrated near the origin).  As a result,
the m.e. of the radiative transitions, $X(3872)\rightarrow J/\psi\gamma$
and $X(3872)\rightarrow \psi(3686)\gamma$, are defined by the ``mixed"
m.e., in which the $2P$ w.f. is calculated with the confining potential,
while the w.f. of $J/\psi$ and $\psi(3686)$ are calculated with the 
linear+GE potential.

These m.e. are denoted as,
\begin{eqnarray}
 K_1({\rm mix}) & = &
 \int dr r^2 R_{2P} ({\rm lin}) |g(r)| R_{1S}({\rm lin+GE}),
\nonumber \\
 K_2({\rm mix}) & = & \int dr r^2~R_{2P}({\rm lin})|g(r)|R_{2S}({\rm lin+GE}).
\label{eq.34}
\end{eqnarray}
It appears that the magnitude of the m.e. $K_1({\rm mix})$ is
larger compared to the number given in Eq.~(\ref{eq.15}) and for the
un-regularized w.f. (the solutions of the SSE in Eq.~(\ref{eq.12}))
the m.e. are
\begin{equation}
 K_1({\rm mix}) = - 0.693~{\rm GeV^{-1}},\quad K_2({\rm mix}) =2.053~{\rm GeV^{-1}}.\label{eq.35}
\end{equation}
Instead, below we use the m.e. $K_i({\rm mix})~(i=1,2)$, calculated for
the regularized w.f.  with the use of the Einbein approximation:
\begin{equation}
 K_1({\rm mix}) = -0.664~{\rm GeV^{-1}},\quad
 K_2({\rm mix}) = 2.475~{\rm GeV^{-1}}.
\label{eq.36}
\end{equation}
For the calculated $K_1({\rm mix})$  the partial width $\Gamma_1$
strongly depends on the value of the mixing parameter $c_{\rm X}$, even
if it is taken from the narrow range given in Eq.~(\ref{eq.30}). Using
the m.e. $I_m$ (which refers to the $1P$ state) from Table~\ref{tab.03}
and $K_1({\rm mix})$ from Eq.~(\ref{eq.36}), one has
\begin{equation}
 \Gamma_1 = \Gamma(X(3872)\rightarrow J/\psi\gamma)= (37\pm 6)
 {\rm~keV~for~} c_{\rm X}=0.200\pm 0.015. 
\label{eq.37}
\end{equation}
Owing to the CC effects the value of $\Gamma_1$ appears to be $\sim
2.2$ times smaller than in SCA. While changing the admissible value of
$c_{\rm X}$ from the minimal to the maximum values, i.e., by $15\%$,
the partial width $\Gamma_1$ increases from 31 keV to 43 keV, i.e., 1.4
times. It means that a precise knowledge of $\Gamma_1$ from experiment
would put a strong restriction on the mixing parameter $c_{\rm X}$
at the resonance energy $ E_{\rm R}=3.872$~GeV.

For the radiative decay $X(3872)\rightarrow \psi(3686)\gamma$ we have
used the radial w.f. $R_{2S}(r)$ for $\psi(3686)$. Then taking the
m.e. $I_{2S,1P}$ from Table~\ref{tab.03} and $K_2({\rm mix})$ from
Eq.~(\ref{eq.36}), one obtains
\begin{equation}
 \Gamma_2=(72\pm 1.0){\rm ~keV~for~} c_{\rm X}=0.200\pm 0.015.
\label{eq.38}
\end{equation}

Its value weakly depends on the mixing parameter $c_{\rm X}$.
However, this width may be smaller, if the $2S-1D$ mixing is
taken into account in the w.f. of $\psi(3686)$ \cite{ref.43}.

Thus in the CC approach with the minimal value of the mixing parameter,
 $c_{\rm X}=0.185$, we have
\begin{equation}
\Gamma_1= 42.7~{\rm keV}, \quad \Gamma_2= 70.5~{\rm keV}, \quad
R_{\psi\gamma} = 1.65,
\label{eq.39}
\end{equation}
while for the maximal value, $c_{\rm X}=0.215$,
\begin{equation}
\Gamma_1=30.8~{\rm keV}, \quad \Gamma_2=73.2~{\rm keV},
 R_{\psi\gamma}=2.38,
\label{eq.40}
\end{equation}
i.e., this partial width $\Gamma_1$ is by $40\%$ smaller than for
$c_{\rm X}(\rm min)=0.185$. The ratio of the widths also changes in
a rather wide range:
\begin{equation}
 R_{\psi\gamma}({\rm th})=2.02\pm 0.36,
\label{eq.41}
\end{equation}
being 2.4 times larger than that in the SCA Eq.~(\ref{eq.20}). This value
is in a good agreement with the LHCb data given in Eq.~(\ref{eq.2}).
Also its lower limit (for $c_{\rm X}=0.185$) does not contradict the
Belle result, $R_{\psi\gamma} < 2.1$.

Thus, in our analysis the agreement with the LHCb result was obtained
under two conditions: when the GE interaction is suppressed and the CC
admixture to the w.f. of $X(3872)$ is not small, $\sim 20\%$. For the
partial width $\Gamma_1$ we predict the value $37\pm 6$~keV.

\section{Conclusions}
\label{sec.VI}

The $c\bar c$ dynamics in $X(3872)$ with $J^{PC}=1^{++}$ was studied
within the CC approach, where a coupling to the $D\bar D^*$ channels
is determined by the string-breaking mechanism and the mixing parameter
$c_{\rm X}(E=E_{\rm R})=0.200\pm 0.015$ was calculated.  Two different
variants of the $c\bar c$ interaction were considered for the higher
charmonium states.
\begin{enumerate}

\item The linear+GE potential used is the same for the lower
and the higher charmonium states. In this case the partial width
$\Gamma_1=\Gamma(X(3872)\rightarrow J/\psi\gamma)\sim 3.0$~keV is small
and gives rise to a very large value of the ratio $R_{\psi\gamma}\gg
1.0$ in the CC approach. On the contrary, in SCA the ratio
$R_{\psi\gamma}=0.84$ is not large.

\item The GE interaction is supposed to be suppressed for higher
states. For this dynamics the charmonium spectrum above the $D\bar
D$ threshold can be described with a good accuracy, if perturbative
corrections in the $c$-quark mass are neglected.  \end{enumerate}
If the suppressed GE potential is used, then the partial width $\Gamma_1$
increases, being very sensitive to the value of the mixing parameter
$ c_{\rm X}$.  For $c_{\rm X}=0.215$ we find  $\Gamma_1= 30.8$~keV,
$\Gamma_2=73.2$~keV, and the calculated ratio of the partial widths
$\rm R_{\psi\gamma}({\rm max})=2.38$  is in good agreement with the LHCb
result. The predicted minimal value, $R_{\psi\gamma}({\rm min})=1.65$,
does not contradict the Belle measurements with $\rm R_{\psi\gamma}(Belle)
< 2.1$.

\begin{acknowledgments}
This work was supported by the grant RFBR 1402-00395.
\end{acknowledgments}

\appendix
\section{The coupled-channel mechanism}
\label{AppendixA}

Here we essentially use the formalism, suggested in Ref.~\cite{ref.44}
and developed further in Refs.~\cite{ref.20,ref.21}, and  below we partly
reiterate the material from Ref.~\cite{ref.21} for the convenience of
the reader.

We use here the string-decay Lagrangian of the ${}^3P_0$ type for the
decay $c\bar c\rightarrow (c\bar q)(\bar c q)$ \cite{ref.20}:

\begin{equation}
 \mathcal{L}_{\rm sd} =\int d^4x~\bar \psi_q M_\omega \psi_q,
\label{eq.B.1}
\end{equation}
where the light quark bispinors are treated in the limit of large $m_c$,
which allows us to go over to the reduced $(2\times 2)$ form of the
decay matrix elements.  Moreover, to simplify the calculations the actual
w.f. of the $c\bar c$ states, calculated in Ref.~\cite{ref.21} with the
use of the RSH, are fitted here by five (or three) oscillator w.f. (HO),
while the $D$ meson w.f. is described by a  single HO term, which provides
a few percent accuracy  with the parameter $\beta_2\simeq 0.48$. In
this case the factor $M_{\omega}$ in (\ref{eq.B.1})  is calculated to be
$M_{\omega}=4\sqrt{2}\sigma/\sqrt{\pi}\beta_2 \simeq 0.8$~GeV (see below),
which produces the  correct total width of $\psi(3770)$. The transition
m.e. for the decays $(c\bar c)_n\rightarrow (D_{n_2}\bar{D}_{n_3})$
are denoted here as $n\rightarrow n_2,n_3$, and in the $2\times 2$
formalism this m.e.  reduces to
\begin{equation}
  J_{nn_2n_3} (\vep) = \frac{\gamma}{\sqrt{N_c}}
 \int \bar{y}_{123} \frac{d^3\veq}{(2\pi)^3} \Psi^+_n
 (\vep +\veq) \psi_{n_2} (\veq) \psi_{n_3} (\veq).
\label{eq.B.2}
\end{equation}
The factor $\bar y_{123}$ contains a trace of the spin-angular
variables (for details see Refs.~\cite{ref.20,ref.21}),
and $\gamma=\frac{2M_{\omega}}{\lan m_q+U_{\rm S} -V_{\rm D}
+\varepsilon_0\ran}$, where the average over the Dirac denominator
contains  the light quark mass $m_q$, the scalar $U_{\rm S}=\sigma r$
and the vector $V_{\rm D}(r)=-\frac{4\alpha}{3 r}$ potentials, and
$\varepsilon_0$, which is the Dirac eigenvalue in the static potential,
created by the heavy quark:
\begin{equation}
 {\vec{\alpha}\cdot\vec{p} +\beta~(m_q + U_{\rm S}(r)})\psi(r) =
 (\epsilon_0 -V_{\rm D}(r))\psi(r)~.
\label{eq.B.3}
\end{equation}
Knowing the string tension $\sigma=0.18$~GeV$^2$ and the averaged
momentum distribution $\beta_2$ in the w.f. of a heavy-light meson,
the value $M_{\omega}=\frac{4\sqrt{2}\sigma}{\sqrt{\pi}\beta_2}$
was calculated in Ref.~\cite{ref.20}. For the $1S$, $2S$, and $3S$
charmonium states $M_{\omega}$ is different and equal to 0.65, 0.80,
1.10, respectively. Using the averaged values of the scalar and vector
potentials, one obtains $\gamma\approx 1.4$.

The intermediate decay channels, like $DD^*$, induce an additional
interaction ``potential" $V_{CC}(\veq,\veq',E)$ (the quotation marks
imply nonlocality and energy dependence of this potential):
\begin{eqnarray}
 V_{\rm CC} (\veq, \veq', E) & = & \sum_{n_2n_3} \int \frac{d^3\vep}{(2\pi)^3}
 \frac{X_{n_2n_3}(\veq, \vep) X_{n_2n_3}^\dagger (\veq', \vep)}{E-E_{n_2n_3} (\vep)}, \\
\label{eq.A.2.4}
 X_{n_2 n_3} (\veq, \vep) & = & \frac{\gamma}{\sqrt{N_c}} \bar
y_{123} (\veq, \vep) \psi_{n_2} (\veq-\vep) \psi_{n_3} (\veq-\vep).
\label{eq.A.2.5}
\end{eqnarray}

The new Hamiltonian with account of the CC interaction is
\begin{equation}
  H=H_0 +V_{\rm CC}
\label{eq.B.6}
\end{equation}
where $H_0$ is given in Eqs.~(\ref{eq.8},\ref{eq.9}).  At this point
it is important to stress that $V_{\rm CC}$ contains all thresholds
of open channels, and to treat the spectrum of the Hamiltonian $H$
rigorously, one should exploit the Weinberg eigenvalue formalism,
developed for this purpose in Ref.~\cite{ref.20}, second reference.
However, for the states below all thresholds (or neglecting the open
channel amplitudes in first approximation) one can use an expansion
in the complete set of the eigenvalues of $H_0$. One can show, as in
Ref.~\cite{ref.20}, that the exact Weinberg formalism reduces to this
expansion for the real eigenvalues at or below thresholds. Using that,
one obtains for the Green function an expansion,
\begin{equation}
 G_{Q\bar Q }(1,2;E) = \sum_{n,m} \psi_n (1)
 (\hat E - E + \hat w)^{-1}_{nm} \psi_m (2),
\label{eq.B.7}
\end{equation}
where the matrix element of $\hat{w}$ is denoted by $w_{nm}$ and given by
\begin{eqnarray}
 w_{nm} (E) & = & \int \frac{d^3\veq}{(2\pi)^3}\frac{d^3\veq'}{(2\pi)^3} \,
 \psi_n (\veq ) V_{\rm CC} (\veq, \veq', E) \psi_m (\veq')
\nonumber \\
 & = &
\int \frac{d^3\vep}{(2\pi)^3}\sum_{n_2n_3} \frac{J_{nn_2n_3}(\vep)
J^\dagger_{mn_2n_3} (\vep)}{E-E_{n_2m_3}(\vep)}
\label{eq.B.8}
\end{eqnarray}
and the energy eigenvalues are obtained from the determinant condition
\begin{equation}
 \det (E-\hat E-\hat w) =0,
\label{eq.B.9}
\end{equation}
where $(\hat{E})_{nm} = E_n \delta_{nm}$

Defining as in Eq.~(\ref{eq.22}) the effective $Q\bar Q$ w.f.,
\begin{equation}
 \psi_{Q\bar Q}^{(\mathcal{B})}=\sum_{k,l} \psi_k \left(
 \frac{1}{\hat E - E + \hat w }\right)_{kl} (\hat B\psi_l) \equiv
\sum_k a_k \psi_k,
\label{eq.B.10}
\end{equation}
one has to calculate the resulting amplitudes $a_k$ in the expansion
of X(3872) in terms of the $Q\bar Q$ states. The values of $w_{nm}$
are given in Table~\ref{tab.04}, and one can see that only two states,
$|1P\ran$ and $|2P\ran$, are important.

Keeping only two $Q\bar Q$  eigenfunctions, one has
\begin{eqnarray}
 a_1 & = & \frac{E_2 -E + w_{22}}{\det} (\hat B \psi_1) -
 \frac{w_{21}}{\det} (\hat B \psi_2) \\
 \label{eq.B.11}
 a_2 & = & \frac{E_1 -E + w_{11}}{\det} (\hat B \psi_2) - \frac{w_{12}}{\det}
(\hat B \psi_1),
\label{eq.B.12}
\end{eqnarray}
where $\det \equiv \det(\hat E - E+ \hat w)$.
Then one obtains the ratio
\begin{equation}
 \frac{a_1}{a_2} = \frac{-w_{21} + (E_2 + w_{22}-E)
 \frac{(\hat B\psi_1)}{(\hat B\psi_2)}}
 {E_1 + w_{11} - E - w_{12}\frac{(\hat B\psi_1)}{(\hat B\psi_2)}}.
\label{eq.B.13}
\end{equation}
We calculate this ratio for the energy $E_{\rm R}=3872$~MeV, taking
the values of $w_{ik}$ from the Table~\ref{tab.04}. Note, that we have
corrected the sign of $w_{12}$ in accordance with the relative signs
of the wave functions $|1P\rangle$ and $|2P\rangle$. The resulting
values of $c_{\rm X}\equiv \frac{a_1}{a_2} (3872$ MeV) are given in
Eqs.~(\ref{eq.28}-\ref{eq.30}).

From Table~\ref{tab.04} one can see that the $3\,{}^3P_1$ state
gives a negligible admixture. The values of $w_{mn}$ for $E=E_R$ are
computed according to Eq.~(\ref{eq.B.8}) with the w.f., calculated in
Ref.~\cite{ref.20}, and are given in Table~\ref{tab.04}.

\begin{table}[h]
\caption{The m.e. $w_{nm}$ (in GeV) between $n\,{}^3P_1$ and $m\,{}^3P_1$
states for two approximations of the exact w.f.}
\label{tab.04}
\begin{tabular}{|c|c|c|c|c|}\hline
   $n$ $m$  &1 1&  1 2& 2 2&3 2\\
\hline
~$w_{nm}$ for 5 HO~&-0.320& 0.122& - 0.099& -0.0003\\
 ~$w_{nm}$ for 3 HO&-0.319& 0.121& - 0.098& -0.0011\\
\hline
\end{tabular}
\end{table}

\end{document}